\documentclass[12pt,english]{article}
\usepackage{lmodern}

\usepackage[T1]{fontenc}
\usepackage[latin9]{inputenc}
\usepackage{geometry}
\usepackage{amsmath}
\usepackage{graphicx}
\usepackage{setspace}
\usepackage{esint}
\usepackage[authoryear]{natbib}
\usepackage{subscript}
\makeatletter

\providecommand{\tabularnewline}{\\}

\newcommand{\lyxaddress}[1]{
	\par {\raggedright #1
	\vspace{1.4em}
	\noindent\par}
}

\@ifundefined{date}{}{\date{}}
\makeatother

\usepackage{babel}
\begin{document}
\title{Cumulative emissions accounting of greenhouse gases due to path independence
for a sufficiently rapid emissions cycle}
\author{Ashwin K Seshadri}
\maketitle

\lyxaddress{Divecha Centre for Climate Change and Centre for Atmospheric and
Oceanic Sciences, Indian Institute of Science, Bangalore 560012, India,
email: ashwin@fastmail.fm}
\begin{abstract}
Cumulative emissions accounting for carbon-dioxide (CO\textsubscript{2})
is founded on recognition that global warming in Earth System Models
(ESMs) is roughly proportional to cumulative CO\textsubscript{2}
emissions, regardless of emissions pathway. However, cumulative emissions
accounting only requires the graph between global warming and cumulative
emissions to be approximately independent of emissions pathway (\textquotedbl path-independence\textquotedbl ),
regardless of functional relationship between these variables. The
concept and mathematics of path-independence are considered for an
energy-balance climate model (EBM), giving rise to a closed-form expression
of global warming, together with analysis of the atmospheric cycle
following emissions. Path-independence depends on the ratio between
the period of the emissions cycle and the atmospheric lifetime, being
a valid approximation if the emissions cycle has period comparable
to or shorter than the atmospheric lifetime. This makes cumulative
emissions accounting potentially relevant beyond CO\textsubscript{2},
to other greenhouse gases (GHGs) with lifetimes of several decades
whose emissions have recently begun. 
\end{abstract}

\section{Introduction}

Several studies have discussed metrics to compare emissions scenarios,
especially where different climate forcers are involved (\citet{Fuglestvedt2003,Shine2005,Allen2016,Frame2019}).
Such a comparison is not easy because the response of the climate
system to radiative forcing is not immediate (\citet{Stouffer2004,Held2010}).
The atmosphere and ocean mixed layer take a few years to reach equilibrium
with forcing (\citet{Held2010,Geoffroy2013,Geoffroy2013a,Seshadri2017}).
Even if this time-delay is neglected, considering the much longer
mitigation policy time-horizons of several decades to centuries, one
still has to reckon with the multi-century timescale of the deep ocean
response to radiative forcing (\citet{Gregory2000,Stouffer2004,Held2010}).
Therefore global warming is not in equilibrium with forcing and it
becomes essential to characterize the non-equilibrium aspect of the
slow-response in order to compare emissions scenarios . Even the simplest
2-box models of global warming yield global warming as a function
of radiative forcing and time, with explicit dependence on the latter
arising due to non-equilibrium effects (\citet{Held2010,Seshadri2017}).
This means that even the limited goal of comparing warming from two
different emissions scenarios from the simplest climate models requires
appealing to uncertain estimates of the relative magnitudes of fast
and slow climate responses, because these depend on different functions
of the emissions graph (\citet{Seshadri2017}). 

For CO\textsubscript{2}, a major simplification resulted from the
finding that its contribution to global warming is proportional to
cumulative emissions (\citet{Allen2009,Matthews2009}), as measured
from preindustrial conditions when emissions are assumed to be negligible.
Several studies have sought to explain this phenomenon, and find that
proportionality arises from approximate cancellation between the concavity
of the radiative forcing relation and the diminishing uptake of heat
and CO\textsubscript{2} into the oceans as global warming proceeds
(\citet{Goodwin2015,MacDougall2015}). Strict proportionality only
occurs for a narrow range of cumulative emissions and elsewhere is
an idealization (\citet{MacDougall2015}), but such a finding of approximate
proportionality across Earth System Models is both surprising and
powerful. For example, it brings about the possibility of ``emergent''
observational constraints on the transient climate response to cumulative
CO\textsubscript{2} emissions (TCRE) and related quantities (\citet{Hall2019,Nijsse2020}),
despite the difficulty of estimating individual parameters constituting
them.

A result of equal importance to mitigation is that different emissions
scenarios of CO\textsubscript{2} can be evaluated by comparing cumulative
emissions alone (\citet{Zickfeld2009,Bowerman2011,Zickfeld2012,Matthews2012,Herrington2014}).
This has served as the foundation for cumulative emissions accounting
in discussions of future mitigation scenarios (\citet{Meinshausen2009,Matthews2012,Stocker2013,Stocker2013a,Millar2017,Friedlingstein2019}),
wherein a certain cumulative emissions budget for CO\textsubscript{2}
leads directly to a distribution of future global warming. According
to the IPCC, \textquotedbl The ratio of GMST {[}global mean surface
temperature{]} change to total cumulative anthropogenic carbon emissions
is relatively constant and independent of the scenario, but is model
dependent, as it is a function of the model cumulative airborne fraction
of carbon and the transient climate response. For any given temperature
target, higher emissions in earlier decades therefore imply lower
emissions by about the same amount later on\textquotedbl{} (\citet{Stocker2013a}).
While the approximately constant ratio between global warming and
cumulative emissions has the implications noted above, in general
cumulative emissions accounting requires only path-independence, and
not necessarily that the ratio be constant. Cumulative emissions accounting
does not require a particular relation: it only requires the graph
between global warming and cumulative emissions to be approximately
independent of emissions pathway (\textquotedbl path-independence\textquotedbl )
(\citet{Zickfeld2012,Herrington2014,Seshadri2017a}). Proportionality
implies path-independence, and the latter is a robust feature of a
wider range of model types, from Earth System Models (ESMs) to much
simpler Energy Balance Models (EBMs). 

Accounting for path-independence is a different type of problem than
accounting for constant TCRE. Accounting for proportionality involves
showing how different effects causing departure from proportionality
cancel each other (\citet{Matthews2009,Goodwin2015,MacDougall2015}).
An explanation must stop here, for it is not possible to explain why
this happens to occur for the Earth system. Contrariwise, explaining
path-independence requires showing how global warming from CO\textsubscript{2}
can be approximated as a function of a single argument, i.e. cumulative
emissions (\citet{Seshadri2017a}). This involves some quantities
being much smaller than others, making counterfactual accounts possible.
In light of this difference, and the sufficiency of path-independence
(not requiring constant TCRE) for cumulative emissions accounting,
this paper considers the mathematics of path-independence in the context
of a two-box EBM, represented by two coupled 1\textsuperscript{st}-order
ordinary differential equations, yielding an explicit formula for
global warming on being integrated.

A simple view of path-independence arises in terms of directional-derivatives
of a function that depends on a few different variables including
cumulative emissions. Radiative forcing can be expressed in terms
of excess CO\textsubscript{2} (compared to preindustrial equilibrium),
which in turn depends on cumulative emissions and airborne fraction
(of cumulative emissions). Hence global warming can be expressed as
a function of cumulative emissions, airborne fraction, and time. During
the increasing phase of cumulative emissions, while net emissions
are positive, path-independence occurs if the increase in global warming
is governed by changes in cumulative emissions, being approximately
the same across scenarios for a given change in cumulative emissions. 

To be concrete, we consider a global warming function $T\left(M,r,t\right)$
where $M$, $r$, and $t$ are cumulative emissions, airborne fraction,
and time. This function is obtained by integrating a simple model
such as an EBM. Along the vector $\overrightarrow{v}=\triangle M\hat{i}+\triangle r\hat{j}+\triangle t\hat{k}$,
with $\triangle$ denoting small changes in these variables with time,
and $\hat{i}$, $\hat{j}$ and $\hat{k}$ denoting unit vectors along
the respective variables' axes, the directional derivative of the
function $T\left(M,r,t\right)$ equals the scalar product (``dot
product'') $\overrightarrow{\nabla}T\circ\overrightarrow{v}$, where
$\overrightarrow{\nabla}T=\frac{\partial T}{\partial M}\hat{i}+\frac{\partial T}{\partial r}\hat{j}+\frac{\partial T}{\partial t}\hat{k}$
is the gradient vector. The formula $\overrightarrow{\nabla}T\circ\overrightarrow{v}$
simply describes the change $\triangle T$ during time-interval $\triangle t$.
For path-independence, this must be approximately equal to the directional
derivative along the axis of cumulative emissions alone, i.e. to $\overrightarrow{\nabla}T\circ\overrightarrow{v}_{M},$where
$\overrightarrow{v}_{M}=\triangle M\hat{i}$. This would result in
the increase in global warming being approximated by effects of changes
in cumulative emissions alone, regardless of differences in airborne
fraction and time between scenarios (\citet{Seshadri2017a}). 

The resulting mathematics is that of inequality constraints, with
some quantities required to be much smaller than others for path-independence
to emerge. In particular, it is required that cumulative emissions
changes much more rapidly, on much shorter timescales, as compared
to the airborne fraction of cumulative emissions and compared to a
derived EBM parameter called the damping timescale, which affects
the magnitude of the slow response and thus the gradient of the function
$T\left(M,r,t\right)$ along the axis of time. The timescale for cumulative
emissions to change depends only on the cycle of emissions, that for
the airborne fraction depends on the atmospheric life-cycle as well
as emissions, whereas the damping timescale is a function of the parameters
of the EBM. Generally, path-independence might arise in scenarios
in any of a few different ways: features of the future emissions pathway,
parameters governing the greenhouse gas's (GHG's) atmospheric cycle,
and counterfactual climate models having quite different damping timescales.
Although the damping timescale is subject to climate modeling uncertainties
and its estimate varies across the two-box EBMs renditions of different
modeling group's ESMs (\citet{Geoffroy2013a,Seshadri2017}), it is
nonetheless least variable among the timescales of interest in the
path independence problem, being constant across climate forcers and
emissions scenarios. 

Therefore this paper considers how path-independence can arise when
the cumulative emissions evolves rapidly compared to the airborne
fraction, assuming that the other constraint involving the effect
of time through the damping timescale is met. For clarity, and since
the case of CO\textsubscript{2} has been examined elsewhere (\citet{Seshadri2017a}),
the present work is focused on climate forcers with a single atmospheric
lifetime. This applies to important greenhouse gases such as nitrous
oxide (N\textsubscript{2}O) and hydrofluorocarbons (HFCs), where
both the atmospheric lifecycle and model of radiative forcing differ
from CO\textsubscript{2} (\citet{Stocker2013a}). Where the path-independence
approximation is valid, emissions scenarios can be approximately compared
directly through cumulative emissions without having to invoke uncertain
model parameters. 

\section{Methods and models}

\subsection{Expression for global warming in a 2-box energy balance model (EBM)}

We consider global warming in a 2-box EBM (\citet{Held2010,Winton2010,Geoffroy2013,Geoffroy2013a}),
consisting of a fast contribution and a slow contribution from deep-ocean
warming that is found to be inversely proportional to a timescale
(``damping timescale'') $\tau_{D}$ (\citet{Seshadri2017}). The
fast contribution is approximately in equilibrium with forcing so
global warming in the EBM is
\begin{equation}
T\left(t\right)\approx\frac{\tau_{f}}{c_{f}}\left(F\left(t\right)+\frac{1}{\tau_{D}}e{}^{-t/\tau_{s}}\int_{0}^{t}e{}^{s/\tau_{s}}F\left(s\right)ds\right)\label{eq:p1}
\end{equation}
with $\tau_{f}$ being the fast time-constant, $\tau_{s}$ the slow
time-constant, $c_{f}$ the heat capacity of the fast system, and
$F\left(t\right)$ is radiative forcing (\citet{Seshadri2017}). The
second term involves integrated effects of deep-ocean warming from
$s=0$ to $s=t$. Symbols appearing in the 2-box EBM are listed in
Table 1.

Defining a new variable proportional to global warming $u\left(t\right)=c_{f}T\left(t\right)/\tau_{f}$,
which has units of radiative forcing, we examine conditions for the
graph of $u\left(t\right)$ versus cumulative emissions $M\left(t\right)$
to be independent of emissions pathway. This would ensure that the
graph of $T\left(t\right)$ versus cumulative emissions $M\left(t\right)$
is also path independent. 

Radiative forcing has species-dependent formula, typically a function
of atmospheric concentration $C\left(t\right)$ at time $t$, which
we denote generally as $F\left(C\left(t\right)\right)$. Excess concentration
in the atmosphere can be written as the product of cumulative emissions
of the species and the airborne fraction of cumulative emissions,
so that concentration becomes $C\left(t\right)=M\left(t\right)r\left(t\right)+C_{eq}$,
where $C_{eq}$ is preindustrial equilibrium concentration. $M\left(t\right)r\left(t\right)$
is the excess concentration, and $M\left(t\right)=\intop_{0}^{t}m\left(s\right)ds$
is cumulative emissions, the integral of emissions $m\left(t\right)$
into the atmosphere. The airborne fraction $r\left(t\right)$ is dimensionless.
Cumulative emissions are counted in the same units as atmospheric
concentrations, describing the concentration increase that would occur
if all of the anthropogenic emissions were to remain in the atmosphere.
These variables are listed in Table 2. 

With these definitions, $u\left(t\right)$ becomes
\begin{alignat}{1}
u\left(t\right) & =F\left(M\left(t\right)r\left(t\right)+C_{eq}\right)+\frac{1}{\tau_{D}}e{}^{-t/\tau_{s}}\int_{0}^{t}e{}^{s/\tau_{s}}F\left(M\left(s\right)r\left(s\right)+C_{eq}\right)ds\label{eq:p2}
\end{alignat}
and depends on cumulative emissions $M\left(t\right)$, airborne fraction
of cumulative emissions $r\left(t\right)$, and time $t$, the latter
appearing explicitly due to the slow climate response involving the
integral term. 

\pagebreak{}

Table 1: Variables and parameters of the 2-box EBM

\begin{tabular}{|c|c|c|}
\hline 
Symbol & Description & Units\tabularnewline
\hline 
\hline 
$T\left(t\right)$ & global warming & K\tabularnewline
\hline 
$\tau_{f}$ & fast time-constant & years\tabularnewline
\hline 
$\tau_{s}$  & slow time-constant & years\tabularnewline
\hline 
$c_{f}$  & heat capacity of the fast system & W years m\textsuperscript{-2} K\textsuperscript{-1}\tabularnewline
\hline 
$\tau_{D}$ & ``damping timescale'' inverse to slow climate response & years\tabularnewline
\hline 
$t$ & time & years\tabularnewline
\hline 
$F\left(t\right)$ & radiative forcing & W m\textsuperscript{-2}\tabularnewline
\hline 
$u\left(t\right)$ & scaled global warming & W m\textsuperscript{-2}\tabularnewline
\hline 
\end{tabular}

\pagebreak{}

Table 2: Emissions and concentration variables and parameters

\begin{tabular}{|c|c|c|}
\hline 
Symbol & Description & Units\tabularnewline
\hline 
\hline 
$m\left(t\right)$ & annual emissions & ppm year\textsuperscript{-1}, Gg year\textsuperscript{-1} etc.\tabularnewline
\hline 
$M\left(t\right)$ & cumulative emissions & ppm , Gg etc.\tabularnewline
\hline 
$C_{eq}$ & preindustrial equilibrium concentration & ppm , Gg etc.\tabularnewline
\hline 
$C\left(t\right)$ & concentration & ppm , Gg etc.\tabularnewline
\hline 
$r\left(t\right)$ & airborne fraction of cumulative emissions & dimensionless\tabularnewline
\hline 
$\tau$ & atmospheric lifetime & years\tabularnewline
\hline 
\end{tabular}

\pagebreak{}

\subsection{Condition for path-independence}

There is path-independence between $u\left(t\right)$ and cumulative
emissions $M\left(t\right)$, with the graph between them being independent
of the emissions pathway, if small changes in $u\left(t\right)$ are
almost entirely accounted for by small changes in $M\left(t\right)$.
Imagining a contour plot of $u\left(t\right)$ versus $M\left(t\right)$,
$r\left(t\right)$, and $t$, a small change in cumulative emissions
allows us to identify the new contour surface of $u$ without regard
to concurrent changes in $r\left(t\right)$ and $t$ in case there
is path-independence. This requires the directional derivative of
$u\left(M,r,t\right)$ along the vector $\overrightarrow{v}=\triangle M\hat{i}+\triangle r\hat{j}+\triangle t\hat{k}$
, which is the dot-product $\overrightarrow{\nabla}u\circ\overrightarrow{v}$,
to be approximated by the directional derivative along the $M$-axis
alone, where $\overrightarrow{\nabla}u=\frac{\partial u}{\partial M}\hat{i}+\frac{\partial u}{\partial r}\hat{j}+\frac{\partial u}{\partial t}\hat{k}$
is the gradient vector. Defining $\overrightarrow{v}_{M}=\triangle M\hat{i}$,
we require approximately $\overrightarrow{\nabla}u\circ\overrightarrow{v}\approx\overrightarrow{\nabla}u\circ\overrightarrow{v}_{M}$,
which leads to the two conditions that the other contributions to
the directional derivative $\overrightarrow{\nabla}u\circ\overrightarrow{v}$
are small in magnitude
\begin{equation}
\left|\frac{\partial u}{\partial r}\triangle r\right|\ll\left|\frac{\partial u}{\partial M}\triangle M\right|\label{eq:p3}
\end{equation}
and
\begin{equation}
\left|\frac{\partial u}{\partial t}\triangle t\right|\ll\left|\frac{\partial u}{\partial M}\triangle M\right|\label{eq:p4}
\end{equation}

It has been shown previously, for the case of CO\textsubscript{2}
, that explicit dependence on time cannot undermine path-independence:
even with a larger slow-response where the left hand side of Eq. (\ref{eq:p4})
would be larger, the right hand side would grow correspondingly as
the slow-response becomes more sensitive to changes in cumulative
emissions (\citet{Seshadri2017a}). Therefore we must compare only
the directional derivatives along axes of $M\left(t\right)$ and $r\left(t\right)$,
and the condition for path independence lies in Eq. (\ref{eq:p3}). 

There is a further complication arising from the slow response. Owing
to the slow response, global warming depends not only on present conditions
but also on past history of radiative forcing. In the present analysis,
this enters through the integral in Eq. (\ref{eq:p2}). The directional
derivative must account for present as well as past changes in the
airborne fraction and cumulative emissions. Therefore the partial
derivatives must operate not only on present $M\left(t\right)$ and
$r\left(t\right)$, but also their past histories, specifically $M\left(s\right)$
and $r\left(s\right)$ for all times $0<s<t$ since the start of anthropogenic
emissions. At time $t$,
\begin{equation}
\frac{\partial u\left(t\right)}{\partial M\left(t\right)}\triangle M\left(t\right)=r\left(t\right)F'\left(M\left(t\right)r\left(t\right)+C_{eq}\right)\triangle M\left(t\right)\label{eq:p5}
\end{equation}
while the cumulative present effect of the uniform perturbation $\triangle M\left(t\right)$\footnote{The choice of perturbation in the conduct of the path independence
analysis is not unique. Other choices could well be made, for example
one might perturb proportionally to $\triangle M\left(s\right)$,
which would then be varying in time and be present in the integral
from $0$ to $t$. Our choice of constant perturbation $\triangle M\left(t\right)$
for all past times $0<s<t$ is motivated by the resulting simplicity
of the analysis and its interpretation. } across $0<s<t$ is
\begin{equation}
\frac{\partial u\left(t\right)}{\partial M\left(s\right)}\triangle M\left(t\right)=\left\{ \frac{1}{\tau_{D}}e{}^{-t/\tau_{s}}\int_{0}^{t}e{}^{s/\tau_{s}}F'\left(M\left(s\right)r\left(s\right)+C_{eq}\right)r\left(s\right)ds\right\} \triangle M\left(t\right)\label{eq:p6}
\end{equation}
Similarly, for the effect of change in airborne fraction at time $t$,
\begin{equation}
\frac{\partial u\left(t\right)}{\partial r\left(t\right)}\triangle r\left(t\right)=M\left(t\right)F'\left(M\left(t\right)r\left(t\right)+C_{eq}\right)\triangle r\left(t\right)\label{eq:p7}
\end{equation}
and the cumulative present effect of the uniform perturbation $\triangle r\left(t\right)$
across $0<s<t$ becomes
\begin{equation}
\frac{\partial u\left(t\right)}{\partial r\left(s\right)}\triangle r\left(t\right)=\left\{ \frac{1}{\tau_{D}}e{}^{-t/\tau_{s}}\int_{0}^{t}e{}^{s/\tau_{s}}F'\left(M\left(s\right)r\left(s\right)+C_{eq}\right)M\left(s\right)ds\right\} \triangle r\left(t\right)\label{eq:p8}
\end{equation}

We recall that the basic condition for path independence is in Eq.
(\ref{eq:p3}). Including the slow response, this becomes
\begin{equation}
\frac{\left|\left\{ \frac{\partial u\left(t\right)}{\partial r\left(t\right)}+\frac{\partial u\left(t\right)}{\partial r\left(s\right)}\right\} \triangle r\left(t\right)\right|}{\left|\left\{ \frac{\partial u\left(t\right)}{\partial M\left(t\right)}+\frac{\partial u\left(t\right)}{\partial M\left(s\right)}\right\} \triangle M\left(t\right)\right|}\ll1\label{eq:p9}
\end{equation}

Upon making 1\textsuperscript{st}-order approximations $\triangle M\left(t\right)\approx\frac{dM\left(t\right)}{dt}\triangle t$
and $\triangle r\left(t\right)\approx\frac{dr\left(t\right)}{dt}\triangle t$,
and describing rates of change in terms of corresponding timescales
\begin{equation}
\frac{dM}{dt}=\frac{1}{\tau_{M}}M\label{eq:p10}
\end{equation}
\begin{equation}
\frac{dr}{dt}=-\frac{1}{\tau_{r}}r\label{eq:p11}
\end{equation}
\footnote{The minus sign in Eq. (\ref{eq:p11}) is because although the airborne
fraction is generally decreasing with time, we would prefer this timescale
to generally take positive values.}this condition for path independence simplifies to
\begin{gather}
\left|\frac{\tau_{M}}{\tau_{r}}\right|\frac{F'\left(M\left(t\right)r\left(t\right)+C_{eq}\right)+\frac{1}{\tau_{D}}e{}^{-t/\tau_{s}}\int_{0}^{t}e{}^{s/\tau_{s}}F'\left(M\left(s\right)r\left(s\right)+C_{eq}\right)\frac{M\left(s\right)}{M\left(t\right)}ds}{F'\left(M\left(t\right)r\left(t\right)+C_{eq}\right)+\frac{1}{\tau_{D}}e{}^{-t/\tau_{s}}\int_{0}^{t}e{}^{s/\tau_{s}}F'\left(M\left(s\right)r\left(s\right)+C_{eq}\right)\frac{r\left(s\right)}{r\left(t\right)}ds}\label{eq:p12}\\
=\left|\frac{\tau_{M}}{\tau_{r}}\right|R_{r}\ll1\label{eq:p13}
\end{gather}
where the second factor is defined as
\begin{equation}
R_{r}\equiv\frac{F'\left(M\left(t\right)r\left(t\right)+C_{eq}\right)+\frac{1}{\tau_{D}}e{}^{-t/\tau_{s}}\int_{0}^{t}e{}^{s/\tau_{s}}F'\left(M\left(s\right)r\left(s\right)+C_{eq}\right)\frac{M\left(s\right)}{M\left(t\right)}ds}{F'\left(M\left(t\right)r\left(t\right)+C_{eq}\right)+\frac{1}{\tau_{D}}e{}^{-t/\tau_{s}}\int_{0}^{t}e{}^{s/\tau_{s}}F'\left(M\left(s\right)r\left(s\right)+C_{eq}\right)\frac{r\left(s\right)}{r\left(t\right)}ds}\label{eq:p14}
\end{equation}

The first terms in both the numerator and denominator of Eq. (\ref{eq:p14})
originate in the sensitivity to present changes in cumulative emissions
and airborne fraction. The second terms involving integrals describe
effects of perturbations to the past values of these variables, through
the slow climate response. 

Moreover since $R_{r}$ is generally on the order of 1, as seen later,
the condition for path independence simplifies to
\begin{equation}
\left|\frac{\tau_{M}}{\tau_{r}}\right|\ll1\label{eq:p15}
\end{equation}

Since this condition in Eq. (\ref{eq:p15}) does not depend on the
integrals, path independence at time $t$ can be understood in terms
of present timescales (at time $t)$ for changes in cumulative emissions
and airborne fraction. This arises independently of the form of the
radiative forcing function $F\left(C\left(t\right)\right)$ or its
derivative $F'\left(C\left(t\right)\right)$. Hence we can examine
path-independence by comparing the atmospheric life-cycle as represented
by $\tau_{r}$ and the cycle of anthropogenic emissions characterized
by $\tau_{M}$. 

Figure 1 illustrates these features for the case of CO\textsubscript{2}.
Global warming is a path-independent function of emissions, because
the directional derivative has a much larger component along the axis
of cumulative emissions, owing to short timescale $\tau_{M}$. Figure
2 shows that the relative effects of airborne fraction and time, compared
to cumulative emissions, as measured by the ratio of directional derivatives,
is approximated well by the ratio of timescales. Moreover, it is bounded
by this ratio, because the second factor in Eq. (\ref{eq:p12}) is
less than 1. Therefore for path-independence it is sufficient that
$\tau_{M}/\tau_{r}$ is small (\citet{Seshadri2017a}). This is generally
the case for other climate forcers as well. 

\begin{figure}
\includegraphics[scale=0.7]{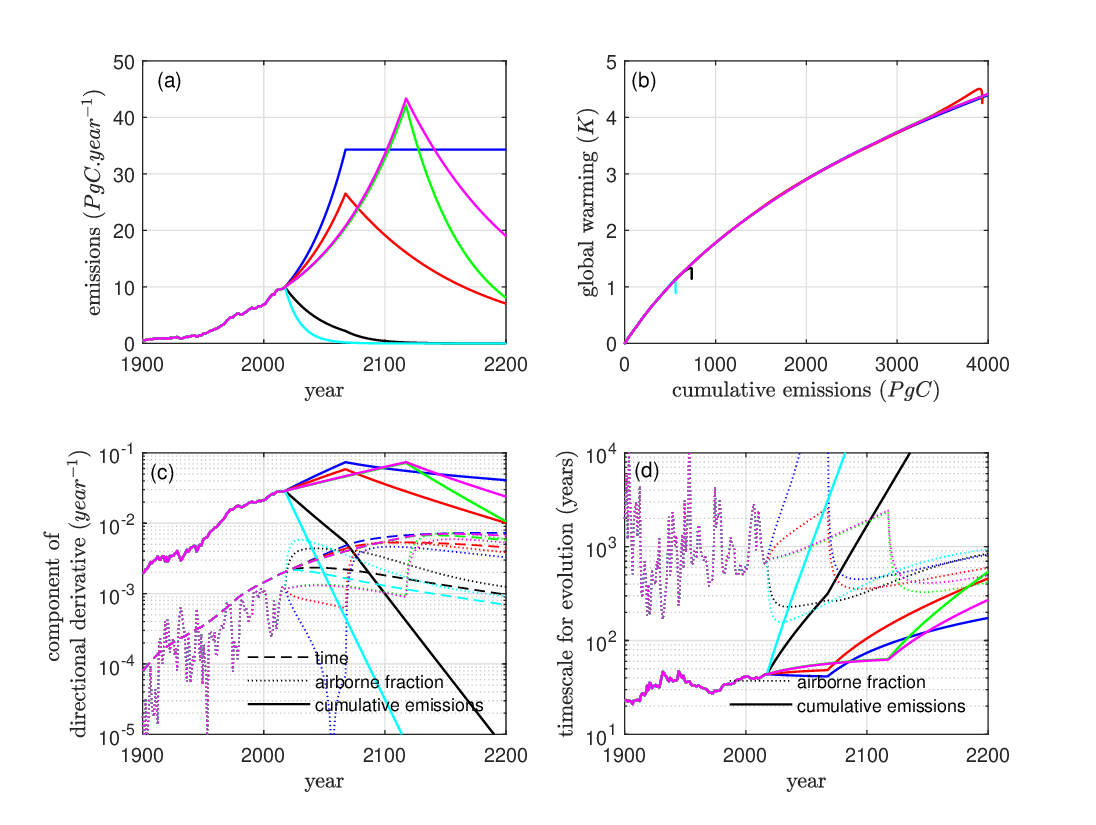}

\caption{Conditions for path-independence between global warming and cumulative
CO\protect\textsubscript{2} emissions in an EBM: (a) emissions scenarios;
(b) global warming versus cumulative CO\protect\textsubscript{2}
emissions, which is path-independent; (c) directional derivatives
of global warming along axes of time, airborne fraction, and cumulative
emissions; (d) timescales for evolution of airborne fraction and cumulative
emissions. Global warming in the EBM is a function of cumulative emissions,
airborne fraction, and time (Eq. (\ref{eq:p1})). It can be approximated
as a function of cumulative emissions alone (panel b) because the
directional derivative along the cumulative emissions axis is much
larger (panel c). This occurs because cumulative emissions evolves
on much shorter timescales (panel d). }
\end{figure}

\begin{figure}
\includegraphics[scale=0.7]{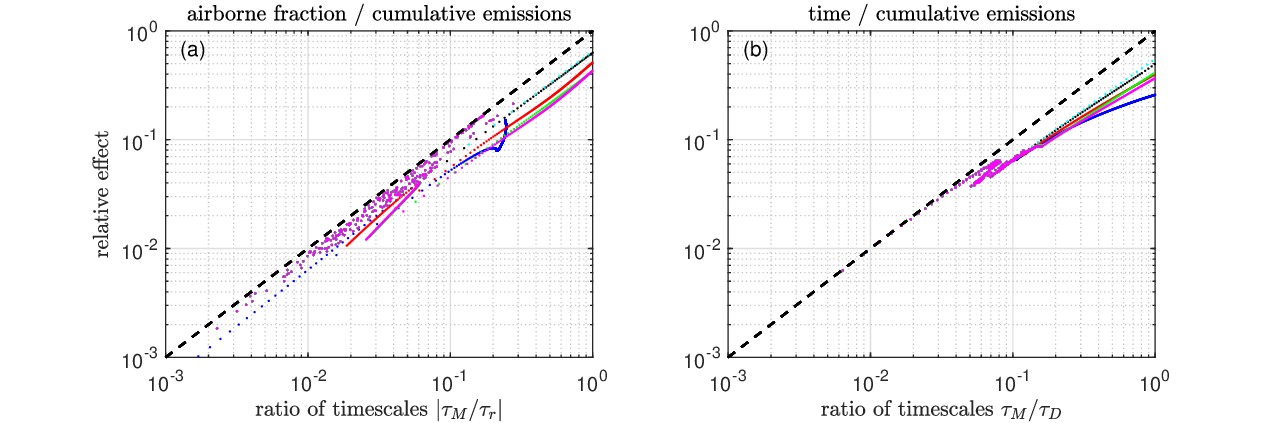}

\caption{The effect of airborne fraction and time on global warming compared
to that of cumulative emissions, as measured by their directional
derivatives: (a) effect of airborne fraction relative to cumulative
emissions versus the ratio of timescales for evolution of cumulative
emissions and airborne fraction, illustrating Eq. (\ref{eq:p13});
(b) effect of time relative to cumulative emissions versus the ratio
of the cumulative emissions timescale and the damping timescale. The
relative effects are bounded by the ratio of timescales, which is
the main influence. For example, the relative effect of changes in
the airborne fraction is small because the airborne fraction evolves
slowly compared to cumulative emissions. Therefore path-independence
requires $\tau_{M}/\tau_{r}$ to be small. }
\end{figure}

\subsection{Model of airborne fraction with a single atmospheric time-constant}

Path-independence for CO\textsubscript{2} with its multiple atmospheric
time-constants was considered in the preceding work (\citet{Seshadri2017a}),
and the present study extends the analysis to develop a model for
$\tau_{M}/\tau_{r}$ in the context of a well-mixed gas with single
atmospheric time-constant $\tau$, which is relevant to many GHGs.
The atmospheric cycle is described by linear differential equation
\begin{equation}
\frac{dC\left(t\right)}{dt}=m\left(t\right)-\frac{C\left(t\right)-C_{eq}}{\tau}\label{eq:p16}
\end{equation}
with emissions $m\left(t\right)$, preindustrial equilibrium concentration
$C_{eq}$ at $t=0$, and atmospheric lifetime $\tau$ with which excess
concentration $C\left(t\right)-C_{eq}$ is reduced. This is integrated
for
\begin{equation}
C\left(t\right)=C_{eq}+e^{-t/\tau}\int_{0}^{t}e^{s/\tau}m\left(s\right)ds\label{eq:p17}
\end{equation}

Writing airborne fraction of cumulative emissions $r\left(t\right)=\left(C\left(t\right)-C_{eq}\right)/M\left(t\right)$,
and differentiating
\begin{equation}
-\frac{1}{\tau_{r}}=\frac{1}{r}\frac{dr}{dt}=\frac{1}{C\left(t\right)-C_{eq}}\frac{d}{dt}\left(C\left(t\right)-C_{eq}\right)-\frac{1}{M\left(t\right)}\frac{d}{dt}M\left(t\right)\label{eq:p18}
\end{equation}
and recognizing $\tau_{M}$ in the last term on the right yields
\begin{equation}
\left|\frac{\tau_{M}}{\tau_{r}}\right|=\left|\frac{\tau_{M}}{C\left(t\right)-C_{eq}}\frac{d}{dt}\left(C\left(t\right)-C_{eq}\right)-1\right|\label{eq:p19}
\end{equation}

which must be small for path-independence. 

\section{Results}

\subsection{Path-independence is more accurate for a short-period emission cycle}

Consider idealized scenarios in which emissions begin at $t=0$, reaching
the maximum value at some intermediate time and subsequently declining
to zero at $t=T$, where $T$ is the period of the emissions cycle
(in years). Cumulative emissions during the cycle is $M\left(t\right)=\int_{0}^{t}m\left(s\right)ds$. 

We introduce a dimensionless time variable $x\equiv t/T$, which ranges
from $0$ to $1$. Emissions is zero at $x=0$, reaches a maximum
at some $0<x<1$ and declines to zero at $x=1$. Introducing this
variable separates the shape of the emissions cycle from its period.
Its shape is defined by shape function $\hat{m}\left(x\right).$ A
specified shape function $\hat{m}\left(x\right)$, defined over its
domain from $0$ to $1,$ describes a family of emissions graphs,
with $m\left(t\right)$ given by the shape function evaluated at the
corresponding $x$, with $m\left(t\right)=\hat{m}\left(x\left(t\right)\right)$.
Each member of the family corresponds to a different value of $T$,
and fixing $T$ fixes $x\left(t\right)=t/T$ . Thus, for a given $\hat{m}\left(x\right)$
emissions cycles differ only in the value of $T.$

We shall consider the case of non-negative emissions, with $\hat{m}\left(x\right)\geq0$
everywhere in the domain. Therefore emissions cycles with longer $T$
evolve more slowly but have larger cumulative emissions by the end
of the emissions cycle. This is seen from the cumulative emissions
equation
\begin{equation}
M\left(t\right)=\int_{0}^{t}m\left(s\right)ds\label{eq:p20}
\end{equation}
where $ds$ is measured in years. We rewrite in terms of $x$ by integrating
between $h=0$ and $h=x$, where $dh$ is now measured in dimensionless
time. In the new variable, cumulative emissions is
\begin{equation}
M\left(t\right)=T\int_{0}^{x\left(t\right)}\hat{m}\left(h\right)dh\label{eq:p21}
\end{equation}
because $ds=Tdh$. Defining a new variable analogous to cumulative
emissions but with the same units as emissions
\begin{equation}
\hat{M}\left(x\right)\equiv\int_{0}^{x}\hat{m}\left(h\right)dh\label{eq:p22}
\end{equation}
which describes the integral of the emissions shape function, now
without regard to period $T$ of the emissions cycle, the cumulative
emissions becomes
\begin{equation}
M\left(t\right)=T\hat{M}\left(x\left(t\right)\right)\label{eq:p23}
\end{equation}

This describes the simple feature that, given a shape function $\hat{m}\left(x\right)$,
cumulative emissions at corresponding positions in the cycle, for
fixed $x$, scales proportionally to the emissions cycle period $T$.
It is useful to examine the cumulative emissions timescale with respect
to the position in the emissions cycle. 

The cumulative emissions timescale, Eq. (\ref{eq:p10}), becomes
\begin{equation}
\tau_{M}\left(t\right)=\frac{M\left(t\right)}{dM\left(t\right)/dt}=\frac{T\hat{M}\left(x\left(t\right)\right)}{T\frac{d\hat{M}\left(x\left(t\right)\right)}{dt}}\label{eq:p24}
\end{equation}
and, recognizing that $\frac{d\hat{M}\left(x\right)}{dt}=\frac{d\hat{M}\left(x\right)}{dx}\frac{dx}{dt}=\frac{1}{T}\frac{d\hat{M}\left(x\right)}{dx}$,
and $\frac{d\hat{M}\left(x\right)}{dx}=\hat{m}\left(x\right)$ from
Eq. (\ref{eq:p22}), we obtain 
\begin{equation}
\tau_{M}\left(t\right)=T\frac{\hat{M}\left(x\left(t\right)\right)}{\hat{m}\left(x\left(t\right)\right)}\label{eq:p25}
\end{equation}

Note that $\hat{M}\left(x\right)$ has the same units as $\hat{m}\left(x\right)$,
one of the advantages in working with this form. For a given emissions
shape function $\hat{m}\left(x\right)$, the cumulative emissions
timescale at any time $t$ is simply $T$ times the ratio $\hat{M}\left(x\right)/$
$\hat{m}\left(x\right)$ determined at the corresponding position
$x\left(t\right)=t/T$ in the emissions cycle. The cumulative emissions
timescale is proportional to the emissions cycle period. Therefore
path-independence, which requires a short cumulative emissions timescale
compared to the airborne fraction timescale, becomes more likely for
a short-period emissions cycle. 

\subsection{Path-independence depends on ratio $T/\tau$}

Moreover, path-independence depends on the ratio $T/\tau$ between
the emissions cycle period and the atmospheric lifetime. We apply
the same rescaling of time as above to demonstrate this. From Eq.
(\ref{eq:p17}) the excess concentration at time $t$ in years is
\begin{equation}
C\left(t\right)-C_{eq}=e^{-t/\tau}\int_{0}^{t}e^{s/\tau}m\left(s\right)ds\label{eq:p26}
\end{equation}
where $ds$ describes increments of time in years. Rewriting the above
integral in terms of $x$ as before,
\begin{equation}
\int_{0}^{t}e^{s/\tau}m\left(s\right)ds=\int_{0}^{x\left(t\right)}e^{hT/\tau}\hat{m}\left(h\right)Tdh\label{eq:p27}
\end{equation}
where, as before, $dh$ describes increments of dimensionless time.
The above follows from $s=hT$, $ds=Tdh$, and emissions at time $s$
being determined by emissions shape function evaluated at corresponding
$h$, i.e. $m\left(s\right)=\hat{m}\left(h\left(s\right)\right)$.
Hence we can rewrite, for excess concentration at $x=t/T$

\begin{equation}
C\left(x\left(t\right)\right)-C_{eq}=Te^{-\alpha x\left(t\right)}\int_{0}^{x\left(t\right)}e^{\alpha h}\hat{m}\left(h\right)dh\label{eq:p28}
\end{equation}

where $\alpha=T/\tau$. Very simply, we have rewritten $t/\tau$ appearing
in Eq. (\ref{eq:p26}) as $\left(T/\tau\right)\left(t/T\right)=\alpha x$,
with the extra $T$ arising from change in integration variable from
$t$ (measured in years) to $x$ (measured in dimensionless time). 

Rewriting the excess concentration in terms of the emissions shape
function and the emissions cycle period illuminates the problem's
structure. Integrating by parts,
\begin{equation}
e^{-\alpha x}\int_{0}^{x}e^{\alpha h}\hat{m}\left(h\right)dh=\hat{m}_{1}\left(x\right)-\alpha\hat{m}_{2}\left(x\right)+\alpha^{2}\hat{m}_{3}\left(x\right)-\ldots\label{eq:p29}
\end{equation}
where $\hat{m}_{i}\left(x\right)=\intop_{0}^{x}\hat{m}_{i-1}\left(h\right)dh$
is the $i$\textsuperscript{th }repeated integral of emissions having
the same units as $\hat{m}\left(x\right)$, and we define $\hat{m}_{0}\left(x\right)\equiv\hat{m}\left(x\right)$.
Then $\hat{M}\left(x\right)$, the integral of the emissions shape
function, becomes $\hat{m}_{1}\left(x\right)$. From Eqs. (\ref{eq:p19})-(\ref{eq:p28})
we obtain the ratio
\begin{equation}
\left|\frac{\tau_{M}}{\tau_{r}}\right|=\left|\frac{1-\alpha\frac{\hat{m}_{1}\left(x\right)}{\hat{m}\left(x\right)}+\alpha^{2}\frac{\hat{m}_{2}\left(x\right)}{\hat{m}\left(x\right)}-\ldots}{1-\alpha\frac{\hat{m}_{2}\left(x\right)}{\hat{m}_{1}\left(x\right)}+\alpha^{2}\frac{\hat{m}_{3}\left(x\right)}{\hat{m}_{1}\left(x\right)}-\ldots}-1\right|\label{eq:p30}
\end{equation}

which must be small for path-independence as shown above. This condition
depends on integrals of the emissions shape function and the value
of dimensionless parameter $\alpha=T/\tau$. For fixed position in
the emissions cycle, i.e. given $x$, accuracy of the path-independence
approximation depends on $\alpha$. In the limit $\alpha\rightarrow0$,
$\left|\frac{\tau_{M}}{\tau_{r}}\right|\rightarrow0$ and path-independence
obviously holds. This is roughly the case of very long atmospheric
lifetime, as compared to the period of the emissions cycle. 

\subsection{Condition for path-independence}

In general $\alpha$ is not necessarily close to zero, even for long-lived
GHGs, but the ratio in Eq. (\ref{eq:p30}) might be small enough that
path-independence is a reasonable approximation. This ratio depends
on $\alpha$ and $x$ alone, once the shape function $\hat{m}\left(x\right)$
is specified. Generally the shape function may be arbitrary, but for
clarity we consider stylized scenarios of the form $\hat{m}\left(x\right)=\beta x^{\gamma}$
during the increasing phase of emissions between $0\leq x<\frac{1}{2}$,
and for the decreasing phase during $\frac{1}{2}\leq x\leq1$, $\hat{m}\left(x\right)=\beta\left(1-x\right)^{\gamma}$,
 a mirror-image. We have assumed $\beta>0$ and $\gamma>0$. Emissions
peaks at $x=\frac{1}{2}$ and declines to zero at $x=1$. We recall
that once the shape function is specified, we must additionally know
the emissions cycle period $T$ to know the emissions profile in time. 

This form of the emissions shape function renders Eq. (\ref{eq:p30})
as a series in $\alpha x$, illuminating the structure of the problem.
During the increasing phase of emissions, repeated integrals are related
as $\hat{m}_{i}\left(x\right)/\hat{m}\left(x\right)=x^{i}/\left\{ \left(\gamma+1\right)\left(\gamma+2\right)\ldots\left(\gamma+i\right)\right\} $
and $\hat{m}_{i+1}\left(x\right)/\hat{m_{1}}\left(x\right)=x^{i}/\left\{ \left(\gamma+2\right)\left(\gamma+3\right)\ldots\left(\gamma+i+1\right)\right\} $
. These terms, appearing in Eq. (\ref{eq:p30}), with increasing powers
of $x<\frac{1}{2}$ and growing factorial-like terms in the denominator,
rapidly decline with $i$. Even for the decreasing phase where $\frac{1}{2}\leq x\leq1$,
these terms decline rapidly and the series converges (as shown in
Supplementary Information, SI). 

For concreteness we consider the increasing phase of emissions, and
stipulate a tolerance $\theta\ll1$ on the accuracy of path-independence,
setting down that path-independence would be an adequate approximation
if $\left|\frac{\tau_{M}}{\tau_{r}}\right|\leq\theta$. In general
the error from neglecting the contributions to the directional derivative
along the axis of $r$ should be quite small, typically much lesser
than 1, for the path independence approximation to be valid. This
requires that in Eq. (\ref{eq:p13}), the product $\theta R_{r}\ll1$.
If we had $R_{r}\ll1$, then $\theta$ would not have to be especially
small. In general, this is not the case, as seen in Figure 2, and
we assume that $\theta$ must be small enough to carry the burden
of limiting the quantity in Eq. (\ref{eq:p13}).

Then the condition for path-independence $0\leq\left|\frac{\tau_{M}}{\tau_{r}}\right|\leq\theta\ll1$
becomes
\begin{multline}
\left|\left\{ 1-\alpha\frac{\hat{m}_{1}\left(x\right)}{\hat{m}\left(x\right)}+\alpha^{2}\frac{\hat{m}_{2}\left(x\right)}{\hat{m}\left(x\right)}-\ldots\right\} -\left\{ 1-\alpha\frac{\hat{m}_{2}\left(x\right)}{\hat{m}_{1}\left(x\right)}+\alpha^{2}\frac{\hat{m}_{3}\left(x\right)}{\hat{m}_{1}\left(x\right)}-\ldots\right\} \right|\\
\leq\theta\left|\left\{ 1-\alpha\frac{\hat{m}_{2}\left(x\right)}{\hat{m}_{1}\left(x\right)}+\alpha^{2}\frac{\hat{m}_{3}\left(x\right)}{\hat{m}_{1}\left(x\right)}-\ldots\right\} \right|\label{eq:p31}
\end{multline}
and, truncating up to $\alpha^{3}$ and defining a new dimensionless
variable $y\equiv\alpha x=t/\tau$, we obtain the condition in terms
of the following inequality in cubic polynomial in $y$
\begin{multline}
g\left(y\right)=\left\{ 3+\theta\left(\gamma+1\right)\right\} y^{3}-\left\{ 2+\theta\left(\gamma+1\right)\right\} \left(\gamma+4\right)y^{2}+\left\{ 1+\theta\left(\gamma+1\right)\right\} \left(\gamma+3\right)\left(\gamma+4\right)y\\
-\theta\left(\gamma+1\right)\left(\gamma+2\right)\left(\gamma+3\right)\left(\gamma+4\right)\leq0\label{eq:p32}
\end{multline}
which is increasing everywhere (please see SI). Furthermore, $g(0)<0$
so that it has positive root $y^{*}$ where $g\left(y^{*}\right)=0$.
This root depends on $\theta$ and $\gamma$. Characterizing the solutions
to this cubic equation is important, and its graph in Figure 3 illustrates
the aforementioned properties.The mathematics of path-independence
reduces to solving this family of cubic polynomials possessing a single
root. The approach illustrates simultaneously three features of the
path independence problem: path independence depends on the value
of $y\equiv\alpha x$, whose effect can in turn be decomposed into
the position in the emissions cycle and the parameter $\alpha$, and
depends on the curvature of the emissions shape function through $\gamma$.

Path-independence with tolerance $\theta$ requires $y<y^{*}\left(\theta,\gamma\right)$,
or $t<\tau y^{*}\left(\theta,\gamma\right)$, and occurs more easily
earlier in the emissions cycle. For example, path-independence is
a valid approximation around the time of the emissions peak at $t=T/2$
if the period of the emissions cycle $T<2\tau y^{*}\left(\theta,\gamma\right)$
and $3/4$\textsuperscript{th} into the emissions cycle if $T<\frac{4}{3}\tau y^{*}\left(\theta,\gamma\right)$,
imposing a progressively more stringent condition further into the
cycle. This is because the cumulative emissions timescale grows with
the position in the cycle, with its value being $\tau_{M}=Tx/\left(\gamma+1\right)$
during the increasing phase of emissions, and a corresponding expression
increasing in $x$ for the decreasing phase of emissions (the latter
is discussed in SI). Symbols used in this section to describe path
independence are listed in Table 3.

\pagebreak{}

Table 3: Symbols used in describing path independence

\begin{tabular}{|c|c|c|}
\hline 
Symbol & Description & Units\tabularnewline
\hline 
\hline 
$\tau_{M}$ & cumulative emissions timescale & years\tabularnewline
\hline 
$\tau_{r}$ & airborne fraction timescale & years\tabularnewline
\hline 
$t$ & time & years\tabularnewline
\hline 
$x$ & rescaled time & dimensionless\tabularnewline
\hline 
$\hat{m}\left(x\right)$ & emissions shape function & same as $m\left(t\right)$ \tabularnewline
\hline 
$\hat{M}\left(x\right)$ & integral of emissions shape function & same as $\hat{m}\left(x\right)$\tabularnewline
\hline 
$\hat{m_{i}}\left(x\right)$ & $i$\textsuperscript{th }repeated integral of $\hat{m}\left(x\right)$ & same as $\hat{m}\left(x\right)$\tabularnewline
\hline 
$\tau$ & atmospheric lifetime & years\tabularnewline
\hline 
$T$ & emissions cycle period & years\tabularnewline
\hline 
$\alpha$ & ratio of $T/\tau$ & dimensionless\tabularnewline
\hline 
$y$ & product $\alpha x=t/\tau$ & dimensionless\tabularnewline
\hline 
$y^{*}$ & root of condition $g\left(y\right)<0$ for path independence & dimensionless\tabularnewline
\hline 
$\gamma$ & exponent of emissions shape function & dimensionless\tabularnewline
\hline 
$\theta$ & tolerance on path-independence accuracy & dimensionless\tabularnewline
\hline 
\end{tabular}\pagebreak{}
\begin{figure}
\includegraphics[scale=0.7]{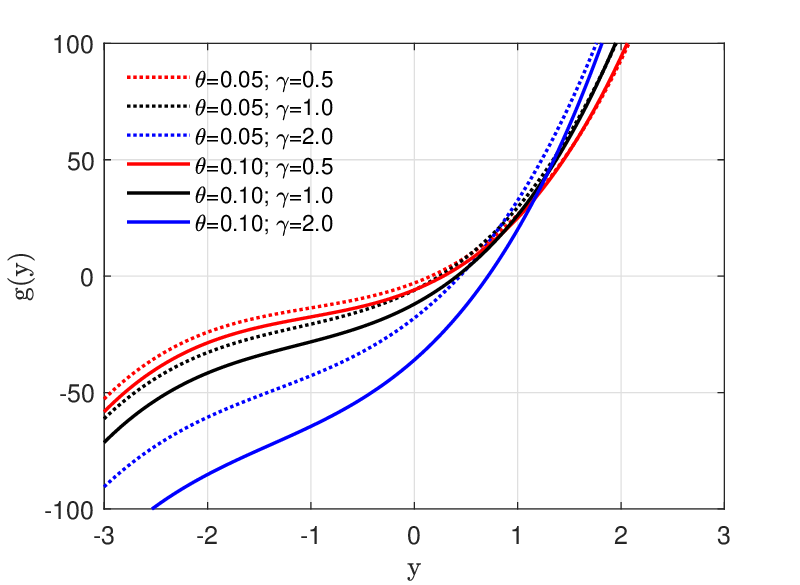}

\caption{Graph of cubic polynomial in Eq. (\ref{eq:p32}). The function is
increasing in $y$, takes a negative value at $y=0$, and has a single
positive root that lies between $0$ and $1$ for parameter values
considered here.}

\end{figure}

In addition to the emissions shape function, it has been assumed that
the 3\textsuperscript{rd}-order truncation of the series in Eq. (\ref{eq:p30})
is adequate. This is justified in Figure 4, which shows contours of
$\theta$ as a function of $x$ and $\alpha$. Owing to the inequality
constraint, path-independence occurs to the lower left side of the
contours. Shown are the exact values, corresponding to the ratio $\left|\frac{\tau_{M}}{\tau_{r}}\right|$
in Eq. (\ref{eq:p19}), and 1\textsuperscript{st}- 3\textsuperscript{rd}
order approximations. The approximation to 3\textsuperscript{rd}
order in $y$ has converged and, since path-independence involves
a constraint on $y=\alpha x$, the resulting boundaries are families
of hyperbolas.

\begin{figure}
\includegraphics[scale=0.7]{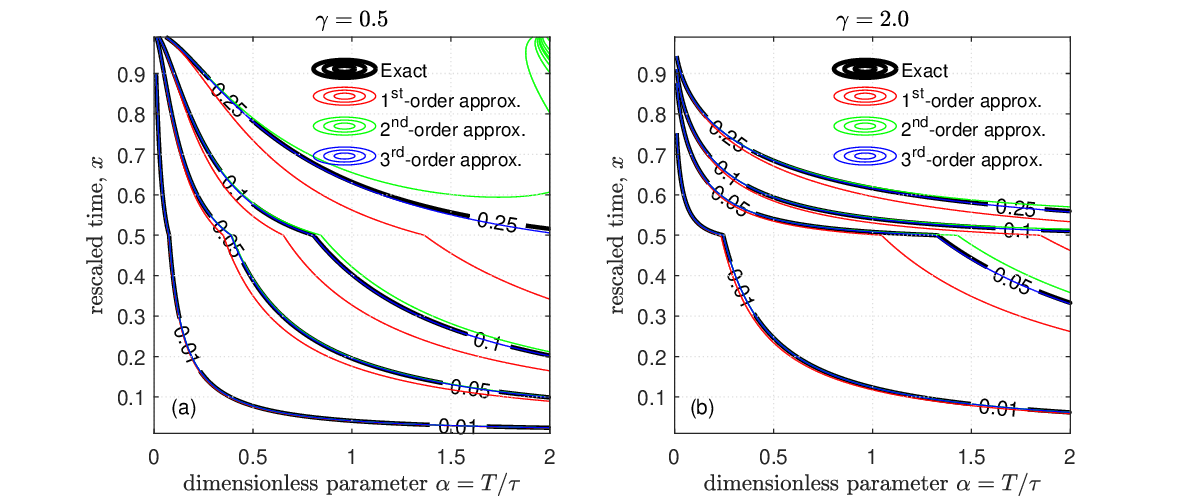}

\caption{Contour plot of tolerance $\theta$ versus the dimensionless parameter
$\alpha=T/\tau$ and rescaled time $x=t/T$. Black lines show exact
values from numerical integration of Eq. (\ref{eq:p19}) whereas colors
show successive series approximations involving Eq. (\ref{eq:p31}).
The cubic approximation performs well for a wide range of the emissions
parameter $\gamma$ and conditions in $\alpha$ and $x$. Therefore
Eq. (\ref{eq:p32}) is an adequate account of path-independence. This
equation provides an inequality constraint on $y=\alpha x$ and the
corresponding boundaries in the figure are hyperbolas. Path-independence
occurs if $\alpha=T/\tau$ is equal to or smaller than the corresponding
contour line. Further into the emissions cycle, measured by higher
$x$, the condition for path-independence becomes more stringent,
requiring smaller $\alpha$. For a more sharply peaked emissions cycle,
involving higher $\gamma,$ the condition for path-independence becomes
less stringent, entailing a rightward shift in the boundaries towards
higher $\alpha$. }

\end{figure}

The shape function, as characterized by $\gamma$, has an important
effect. The cumulative emissions timescale during the increasing phase
is $\tau_{M}=Tx/\left(\gamma+1\right)$ as noted earlier, which is
short for small $T$ or large $\gamma$. This is also true of the
decreasing phase (please refer to SI). Hence more sharply-peaked emissions
graphs, with higher $\gamma$, can lead to shorter cumulative emissions
timescales and satisfy path-independence to a better approximation.
The main influence remains of course the period of the emissions cycle.
These results are summarized in Figure 5, which shows that the ratio
of timescales $\left|\frac{\tau_{M}}{\tau_{r}}\right|$ is a function
of $\alpha=T/\tau$. The figure also depicts the range of validity
of the cubic approximation, which fails for higher $\alpha$. 

\begin{figure}
\includegraphics[scale=0.7]{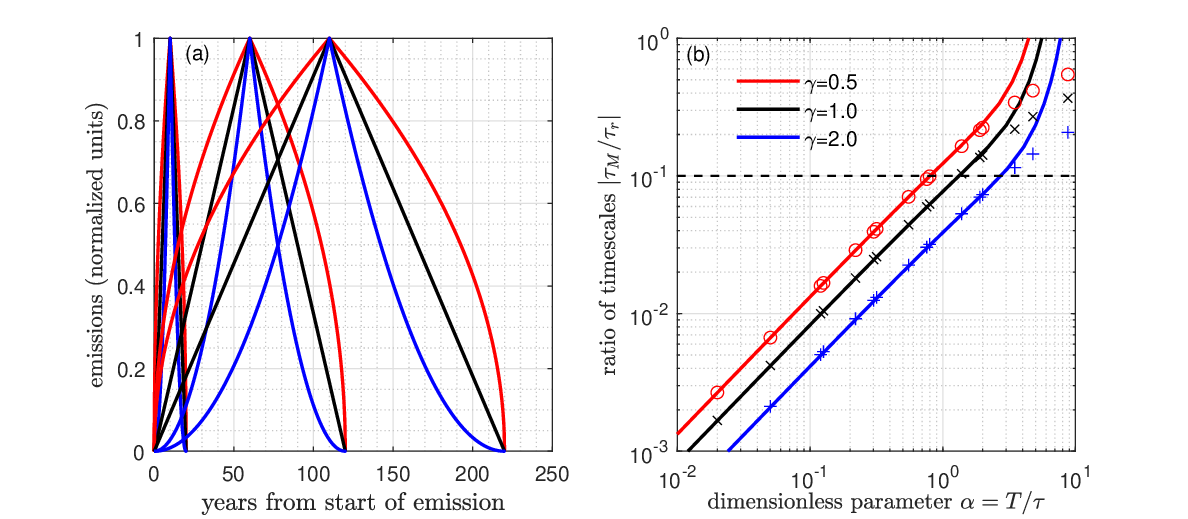}

\caption{The ratio of timescales $\left|\tau_{M}/\tau_{r}\right|$ at the peak
of the emissions graph, i.e. $x=\frac{1}{2}$, as a function of the
dimensionless parameter $\alpha=T/\tau$ and the sharpness of the
emissions shape function, as measured by exponent $\gamma$. This
is drawn for different emissions graphs depicted in the left panel.
On the right panel, markers describe exact values from Eq. (\ref{eq:p19}),
whereas the continuous lines show approximations from the model of
Eq. (\ref{eq:p30}) truncated to cubic terms. The model is valid for
small ratio of timescales, including where $\left|\tau_{M}/\tau_{r}\right|\protect\leq0.1$
for which it is stipulated that path-independence holds. This value
of $\left|\tau_{M}/\tau_{r}\right|$ is governed by the ratio between
the period of the emissions cycle and the atmospheric lifetime. For
example, with $\gamma$ between $0.5$ and $1.0$, the ratio $\left|\tau_{M}/\tau_{r}\right|<0.1$
at the emissions peak if, approximately, $T\protect\leq\tau$. For
a given atmospheric lifetime, this entails a short emissions cycle.
For a more sharply peaked emissions cycle with higher $\gamma,$ path-independence
can occur for larger $T$, corresponding to families of emissions
cycles having longer periods. }
\end{figure}

\subsection{Example}

The above account of path-independence has practical relevance even
to those GHGs with lifetime considerably shorter than CO\textsubscript{2},
whose emissions have recently begun. A good example is hydrofluorocarbons
(HFC) that replaced ozone-depleting substances under the implementation
of the Montreal Protocol, with their emissions beginning to grow during
the early 1990s (\citet{Velders2009,Lunt2015,Stanley2020}). These
HFCs are greenhouse gases with large global warming potentials, and
individual lifetimes ranging from few years to a few hundred years
(\citet{Naik2000}). Their present contribution to radiative forcing
is quite small, but projected to increase by two orders of magnitude
by end of century, to larger than 0.1 W m\textsuperscript{-2}, if
unabated (\citet{Velders2009,Zhang2011,Velders2015}). In the future
some important contributors are expected to include HFC22 (12), HFC134a
(14), HFC125 (29), HFC152a (1.4), HFC143a (52), and HFC32 (4.9), with
estimated lifetimes (in years) listed in brackets (\citet{Naik2000,Zhang2011}). 

\begin{figure}
\includegraphics[scale=0.6]{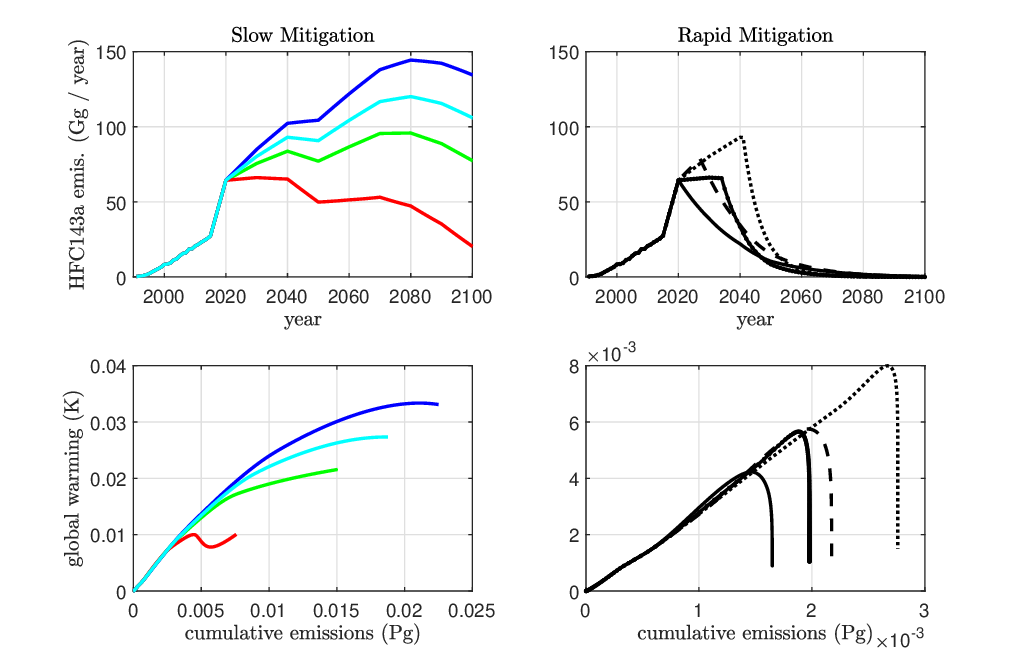}

\caption{Global warming versus cumulative emissions for HFC143a, with atmospheric
lifetime of 52 years. Left-side panels show scenarios involving slow
mitigation, whereas right-side panels show rapid mitigation. Each
color in the left-side panels shows a different emissions scenario,
and each linestyle in the right-side panels shows a different scenario.
Global warming is independent of emissions pathway if the emissions
cycles play out over timescales comparable to the atmospheric lifetime
(right-side panels). }
\end{figure}

The example of HFC143a (Figure 6) is particularly relevant, because
its lifetime of 52 years is still longer than the roughly 30-year
period that has elapsed since its emissions began (\citet{Orkin1996,Zhang2011}).
Past annual emissions are drawn from \citet{Simmonds2017}, for the
period 1991-2015. During the years 2016 to 2019, we interpolate the
RCP3 scenario (\citet{Vuuren2007,Meinshausen2011a}) and future scenarios
in the left panels are modification of the basic RCP3 scenario, with
both larger and smaller emissions scenarios included (Figure 6a).
Radiative forcing of HFC143a is linear in concentrations (in ppbv)
(\citet{Naik2000,Zhang2011,Stocker2013a}). Global warming is computed
for HFC143a from numerical integrations of the 2-box EBM (\citet{Held2010}).
The lower-left panel (c) shows that cumulative emissions accounting
is not applicable for these scenarios because the relation between
global warming and cumulative emissions is not path-independent. The
right panel (Figure 6b) presents scenarios involving much more rapid
mitigation, so that emissions decrease to nearly zero by the middle
of the 21\textsuperscript{st} century. For these scenarios, during
the growing phase of cumulative emissions of HFC143a while annual
emissions are non-zero, global warming is approximately a function
of cumulative emissions. This breaks down only after cumulative emissions
stop growing and the global warming contribution begins to decrease
following a draw-down of concentrations, as also occurs for CO\textsubscript{2}.
During the increasing phase of cumulative emissions, path-independence
is a good approximation, supporting cumulative emissions accounting
for such a scenario family. 

\section{Discussion}

Path-independence of the relation between global warming and cumulative
emissions is adequate for cumulative emissions accounting, and hence
its mathematics bears examination. The question of path-independence
is whether global warming can be approximated by cumulative emissions
alone, and this occurs where directional derivatives of global warming
with respect to the other parameters are small compared to the effect
of cumulative emissions. The idealized account of this paper, based
on an explicit formula for global warming obtained by integrating
an EBM (\citet{Seshadri2017}), shows that no special model of radiative
forcing or atmospheric cycle is necessary. Path-independence only
requires the emissions cycle to progress rapidly enough that the cumulative
emissions timescale is short compared to the timescale for evolution
of the airborne fraction. This broadens the potential relevance of
cumulative emissions accounting beyond CO\textsubscript{2}, especially
to other GHGs with lifetimes of several decades whose emissions have
recently begun. 

The cumulative emissions timescale depends on the period of the emissions
cycle and its shape, but not its amplitude. Quantitatively, if the
emissions cycle proceeds with a period comparable to or less than
the atmospheric lifetime, path-independence is approximately valid,
because the effects of changes in airborne fraction are an order of
magnitude smaller than the effect of growing cumulative emissions.
In effect, if the emissions cycle is rapid compared to the atmospheric
lifetime, it does not matter when the species is emitted. 

For GHGs with lifetime on the range of several decades, this could
occur if the emissions cycle proceeds rapidly enough. This has been
illustrated for HFC143a. Although this has a much shorter atmospheric
lifetime than CO\textsubscript{2} of about 52 years, since its emissions
began only in the 1990s (\citet{Simmonds2017}), cumulative emissions
accounting would be applicable to this species while considering scenarios
where the emissions were to be reduced substantially if not eliminated
during the first half of the 21\textsuperscript{st} century. Of course,
this is not relevant to the much shorter-lived HFCs, or to short-lived
climate pollutants in general. 

Path-independence is a straightforward result of differing timescales,
and does not depend on particular physics of climate forcers or the
Earth system. The present model considers only single atmospheric
time-constants, but can be readily extended to multiple time-constants
or baskets of greenhouse gases such as the HFCs considered together
(\citet{Velders2009,Velders2015}). It makes predictions that would
be useful to examine using more complex models. Sharply-peaked emissions
cycles present rapid evolution of cumulative emissions, with shorter
cumulative emissions timescales, so path-independence can occur somewhat
more easily in this case. The adequacy of the path-independence approximation
degrades with progression of the emissions cycle, owing to increase
in the cumulative emissions timescale with time. These are made explicit
using the power-law model of the emissions cycle, but the qualitative
conclusions are broadly applicable and observed in the results involving
the more general emissions scenarios presented here. 

\section*{Acknowledgments}

The author thanks Prof. Govindasamy Bala, and seminar participants
at Imperial College, London, and University of Exeter for helpful
discussions. 

\bibliographystyle{agufull08}
\bibliography{Refs}

\begin{thebibliography}{38}
\providecommand{\natexlab}[1]{#1}
\expandafter\ifx\csname urlstyle\endcsname\relax
  \providecommand{\doi}[1]{doi:\discretionary{}{}{}#1}\else
  \providecommand{\doi}{doi:\discretionary{}{}{}\begingroup
  \urlstyle{rm}\Url}\fi

\bibitem[{\textit{Allen et~al.}(2009)\textit{Allen, Frame, Huntingford, Jones,
  Lowe, Meinshausen, and Meinshausen}}]{Allen2009}
Allen, M.~R., D.~J. Frame, C.~Huntingford, C.~D. Jones, J.~A. Lowe,
  M.~Meinshausen, and N.~Meinshausen (2009), Warming caused by cumulative
  carbon emissions towards the trillionth tonne, \textit{Nature}, \textit{458},
  1163--1166, \doi{http://dx.doi.org/10.1038/nature08019}.

\bibitem[{\textit{Allen et~al.}(2016)\textit{Allen, Fuglestvedt, Shine,
  Reisinger, Pierrehumbert, and Forster}}]{Allen2016}
Allen, M.~R., J.~S. Fuglestvedt, K.~P. Shine, A.~Reisinger, R.~T.
  Pierrehumbert, and P.~M. Forster (2016), New use of global warming potentials
  to compare cumulative and short-lived climate pollutants, \textit{Nature
  Climate Change}, \textit{6}, 773--777, \doi{10.1038/NCLIMATE2998}.

\bibitem[{\textit{Bowerman et~al.}(2011)\textit{Bowerman, Frame, Huntingford,
  Lowe, and Allen}}]{Bowerman2011}
Bowerman, N. H.~A., D.~J. Frame, C.~Huntingford, J.~A. Lowe, and M.~R. Allen
  (2011), Cumulative carbon emissions, emissions floors and short-term rates of
  warming: implications for policy, \textit{Philosophical Transactions of the
  Royal Society of London A}, \textit{369}, 45--66,
  \doi{http://dx.doi.org/10.1098/rsta.2010.0288}.

\bibitem[{\textit{Frame et~al.}(2019)\textit{Frame, Harrington, Fuglestvedt,
  Millar, Joshi, and Caney}}]{Frame2019}
Frame, D.~J., L.~J. Harrington, J.~S. Fuglestvedt, R.~J. Millar, M.~M. Joshi,
  and S.~Caney (2019), Emissions and emergence: a new index comparing relative
  contributions to climate change with relative climatic consequences,
  \textit{Environmental Research Letters}, \textit{14}, 1--10,
  \doi{10.1088/1748-9326/ab27fc}.

\bibitem[{\textit{Friedlingstein et~al.}(2019)\textit{Friedlingstein, Jones,
  O'Sullivan, Andrew, Hauck, and Peters}}]{Friedlingstein2019}
Friedlingstein, P., M.~W. Jones, M.~O'Sullivan, R.~M. Andrew, J.~Hauck, and
  G.~P. Peters (2019), {Global Carbon Budget} 2019, \textit{Earth System
  Science Data}, \textit{11}, 1783--1838, \doi{10.5194/essd-11-1783-2019}.

\bibitem[{\textit{Fuglestvedt et~al.}(2003)\textit{Fuglestvedt, Berntsen,
  Godal, Sausen, Shine, and Skodvin}}]{Fuglestvedt2003}
Fuglestvedt, J.~S., T.~K. Berntsen, O.~Godal, R.~Sausen, K.~P. Shine, and
  T.~Skodvin (2003), Metrics of climate change: assessing radiative forcing and
  emission indices, \textit{Climatic Change}, \textit{58}, 267--331,
  \doi{10.1023/A:1023905326842}.

\bibitem[{\textit{Geoffroy et~al.}(2013{\natexlab{a}})\textit{Geoffroy,
  Saint-Martin, Olivie, Voldoire, Bellon, and Tyteca}}]{Geoffroy2013}
Geoffroy, O., D.~Saint-Martin, D.~J.~L. Olivie, A.~Voldoire, G.~Bellon, and
  S.~Tyteca (2013{\natexlab{a}}), Transient climate response in a two-layer
  energy-balance model: Part 1: analytical solution and parameter calibration
  using {C}{M}{I}{P}5 {A}{O}{G}{C}{M} experiments, \textit{Journal of Climate},
  \textit{26}, 1841--1857, \doi{http://dx.doi.org/10.1175/JCLI-D-12-00195.1}.

\bibitem[{\textit{Geoffroy et~al.}(2013{\natexlab{b}})\textit{Geoffroy,
  Saint-Martin, Bellon, Voldoire, Olivie, and Tyteca}}]{Geoffroy2013a}
Geoffroy, O., D.~Saint-Martin, G.~Bellon, A.~Voldoire, D.~J.~L. Olivie, and
  S.~Tyteca (2013{\natexlab{b}}), Transient climate response in a two-layer
  energy-balance model. {P}art {I}{I}: {R}epresentation of the efficacy of
  deep-ocean heat uptake and validation for {C}{M}{I}{P}5 {A}{O}{G}{C}{M}s,
  \textit{Journal of Climate}, \textit{26}, 1859--1876,
  \doi{http://dx.doi.org/10.1175/JCLI-D-12-00196.1}.

\bibitem[{\textit{Goodwin et~al.}(2015)\textit{Goodwin, Williams, and
  Ridgwell}}]{Goodwin2015}
Goodwin, P., R.~G. Williams, and A.~Ridgwell (2015), Sensitivity of climate to
  cumulative carbon emissions due to compensation of ocean heat and carbon
  uptake, \textit{Nature Geoscience}, \textit{8}, 29--34,
  \doi{http://dx.doi.org/10.1038/ngeo2304}.

\bibitem[{\textit{Gregory}(2000)}]{Gregory2000}
Gregory, J.~M. (2000), Vertical heat transports in the ocean and their effect
  on time-dependent climate change, \textit{Climate Dynamics}, \textit{16},
  501--515, \doi{http://dx.doi.org/10.1007/s003820000059}.

\bibitem[{\textit{Hall et~al.}(2019)\textit{Hall, Cox, Huntingford, and
  Klein}}]{Hall2019}
Hall, A., P.~Cox, C.~Huntingford, and S.~Klein (2019), Progressing emergent
  constraints on future climate change, \textit{Nature Climate Change},
  \textit{9}, 269--278, \doi{10.1038/s41558-019-0436-6}.

\bibitem[{\textit{Held et~al.}(2010)\textit{Held, Winton, Takahashi, Delworth,
  Zeng, and Vallis}}]{Held2010}
Held, I.~M., M.~Winton, K.~Takahashi, T.~Delworth, F.~Zeng, and G.~K. Vallis
  (2010), Probing the fast and slow components of global warming by returning
  abruptly to preindustrial forcing, \textit{Journal of Climate}, \textit{23},
  2418--2427, \doi{http://dx.doi.org/10.1175/2009JCLI3466.1}.

\bibitem[{\textit{Herrington and Zickfeld}(2014)}]{Herrington2014}
Herrington, T., and K.~Zickfeld (2014), Path independence of climate and carbon
  cycle response over a broad range of cumulative carbon emissions,
  \textit{Earth System Dynamics}, \textit{5}, 409--422,
  \doi{http://dx.doi.org/10.5194/esd-5-409-2014}.

\bibitem[{\textit{Lunt et~al.}(2015)\textit{Lunt, Rigby, Ganesan, Manning,
  Prinn, and O'Doherty}}]{Lunt2015}
Lunt, M.~F., M.~Rigby, A.~L. Ganesan, A.~J. Manning, R.~G. Prinn, and
  S.~O'Doherty (2015), {Reconciling reported and unreported HFC emissions with
  atmospheric observations}, \textit{Proceedings of the National Academy of
  Sciences}, \textit{112}, 5927--5931, \doi{10.1073/pnas.1420247112}.

\bibitem[{\textit{MacDougall and Friedlingstein}(2015)}]{MacDougall2015}
MacDougall, A.~H., and P.~Friedlingstein (2015), The origin and limits of the
  near proportionality between climate warming and cumulative {C}{O}2
  emissions, \textit{Journal of Climate}, \textit{28}, 4217--4230,
  \doi{http://dx.doi.org/10.1175/JCLI-D-14-00036.1}.

\bibitem[{\textit{Matthews et~al.}(2009)\textit{Matthews, Gillett, Stott, and
  Zickfeld}}]{Matthews2009}
Matthews, H.~D., N.~P. Gillett, P.~A. Stott, and K.~Zickfeld (2009), The
  proportionality of global warming to cumulative carbon emissions,
  \textit{Nature}, \textit{459}, 829--832,
  \doi{http://dx.doi.org/10.1038/nature08047}.

\bibitem[{\textit{Matthews et~al.}(2012)\textit{Matthews, Solomon, and
  Pierrehumbert}}]{Matthews2012}
Matthews, H.~D., S.~Solomon, and R.~T. Pierrehumbert (2012), Cumulative carbon
  as a policy framework for achieving climate stabilization,
  \textit{Philosophical Transactions of the Royal Society of London A},
  \textit{370}, 4365--4379, \doi{http://dx.doi.org/10.1098/rsta.2012.0064}.

\bibitem[{\textit{Meinshausen et~al.}(2009)\textit{Meinshausen, Meinshausen,
  Hare, Raper, Frieler, Knutti, Frame, and Allen}}]{Meinshausen2009}
Meinshausen, M., N.~Meinshausen, W.~Hare, S.~C.~B. Raper, K.~Frieler,
  R.~Knutti, D.~J. Frame, and M.~R. Allen (2009), Greenhouse-gas emission
  targets for limiting global warming to 2{C}, \textit{Nature (London, United
  Kingdom)}, \textit{458}, 1158--1162, \doi{10.1038/nature08017}.

\bibitem[{\textit{Meinshausen et~al.}(2011)\textit{Meinshausen, Smith, Calvin,
  Daniel, and Kainuma}}]{Meinshausen2011a}
Meinshausen, M., S.~J. Smith, K.~Calvin, J.~S. Daniel, and M.~L.~T. Kainuma
  (2011), The {R}{C}{P} greenhouse gas concentrations and their extensions from
  1765 to 2300, \textit{Climatic Change}, \textit{109}, 213--241,
  \doi{http://dx.doi.org/10.1007/s10584-011-0156-z}.

\bibitem[{\textit{Millar et~al.}(2017)\textit{Millar, Fuglestvedt,
  Friedlingstein, Grubb, J., Rogelj, and Matthews}}]{Millar2017}
Millar, R.~J., J.~S. Fuglestvedt, P.~Friedlingstein, M.~Grubb, J., Rogelj, and
  H.~D. Matthews (2017), {Emission budgets and pathways consistent with
  limiting warming to 1.5°C}, \textit{Nature Geoscience}, \textit{10},
  741--747, \doi{10.1038/ngeo3031}.

\bibitem[{\textit{Naik et~al.}(2000)\textit{Naik, Jain, Patten, and
  Wuebbles}}]{Naik2000}
Naik, V., A.~K. Jain, K.~O. Patten, and D.~J. Wuebbles (2000), {Consistent sets
  of atmospheric lifetimes and radiative forcings on climate for CFC
  replacements: HCFCs and HFCs}, \textit{Journal of Geophysical Research},
  \textit{105}, 6903--6914, \doi{10.1029/1999JD901128}.

\bibitem[{\textit{Nijsse et~al.}(2020)\textit{Nijsse, Cox, and
  Williamson}}]{Nijsse2020}
Nijsse, F.~J., P.~M. Cox, and M.~S. Williamson (2020), {An emergent constraint
  on Transient Climate Response from simulated historical warming in CMIP6
  models}, \textit{Earth System Dynamics Discussions}, pp. 1--14,
  \doi{10.5194/esd-2019-86}.

\bibitem[{\textit{Orkin et~al.}(1996)\textit{Orkin, Huie, and
  Kurylo}}]{Orkin1996}
Orkin, V.~L., R.~E. Huie, and M.~J. Kurylo (1996), {Atmospheric Lifetimes of
  HFC-143a and HFC-245fa: Flash Photolysis Resonance Fluorescence Measurements
  of the OH Reaction Rate Constants}, \textit{The Journal of Physical
  Chemistry}, \textit{100}, 8907--8912, \doi{10.1021/jp9601882}.

\bibitem[{\textit{Seshadri}(2017{\natexlab{a}})}]{Seshadri2017}
Seshadri, A.~K. (2017{\natexlab{a}}), Fast-slow climate dynamics and peak
  global warming, \textit{Climate Dynamics}, \textit{48}, 2235--2253,
  \doi{10.1007/s00382-016-3202-8}.

\bibitem[{\textit{Seshadri}(2017{\natexlab{b}})}]{Seshadri2017a}
Seshadri, A.~K. (2017{\natexlab{b}}), {Origin of path independence between
  cumulative CO2 emissions and global warming}, \textit{Climate Dynamics},
  \textit{49}, 3383--3401, \doi{10.1007/s00382-016-3519-3}.

\bibitem[{\textit{Shine et~al.}(2005)\textit{Shine, Fuglestvedt, Hailemariam,
  and Stuber}}]{Shine2005}
Shine, K., J.~S. Fuglestvedt, K.~Hailemariam, and N.~Stuber (2005),
  Alternatives to the global warming potential for comparing climate impacts of
  emissions of greenhouse gases, \textit{Climatic Change}, \textit{68},
  281--302, \doi{10.1007/s10584-005-1146-9}.

\bibitem[{\textit{Simmonds et~al.}(2017)\textit{Simmonds, Rigby, McCulloch,
  O'Doherty, Young, and Muhle}}]{Simmonds2017}
Simmonds, P.~G., M.~Rigby, A.~McCulloch, S.~O'Doherty, D.~Young, and J.~Muhle
  (2017), {Changing trends and emissions of hydrochlorofluorocarbons (HCFCs)
  and their hydrofluorocarbon (HFCs) replacements}, \textit{Atmospheric
  Chemistry and Physics}, \textit{17}, 4641--4655,
  \doi{10.5194/acp-17-4641-2017}.

\bibitem[{\textit{Stanley et~al.}(2020)\textit{Stanley, Say, Muhle, Harth,
  Krummel, Young, and O'Doherty}}]{Stanley2020}
Stanley, K., D.~Say, J.~Muhle, C.~Harth, P.~Krummel, D.~Young, and S.~O'Doherty
  (2020), {Increase in global emissions of HFC-23 despite near-total expected
  reductions}, \textit{Nature Communications}, \textit{396}, 1--6,
  \doi{10.1038/s41467-019-13899-4}.

\bibitem[{\textit{Stocker}(2013)}]{Stocker2013}
Stocker, T.~F. (2013), The closing door of climate targets, \textit{Science},
  \textit{339}, 280--282, \doi{http://dx.doi.org/10.1126/science.1232468}.

\bibitem[{\textit{Stocker et~al.}(2013)\textit{Stocker, Qin, Plattner,
  Alexander, and Allen}}]{Stocker2013a}
Stocker, T.~F., D.~Qin, G.-K. Plattner, L.~V. Alexander, and S.~K. Allen
  (2013), \textit{Climate Change 2013: The Physical Science Basis. Contribution
  of Working Group I to the Fifth Assessment Report of the Intergovernmental
  Panel on Climate Change}, chap. Technical Summary, pp. 33--118, Cambridge
  University Press.

\bibitem[{\textit{Stouffer}(2004)}]{Stouffer2004}
Stouffer, R.~J. (2004), Time scales of climate response, \textit{Journal of
  Climate}, \textit{17}, 209--217,
  \doi{http://dx.doi.org/10.1175/1520-0442(2004)017<0209:TSOCR>2.0.CO;2}.

\bibitem[{\textit{van Vuuren et~al.}(2007)\textit{van Vuuren, den Elzen, Lucas,
  Eickhout, and Strengers}}]{Vuuren2007}
van Vuuren, D.~P., M.~G.~J. den Elzen, P.~L. Lucas, B.~Eickhout, and B.~J.
  Strengers (2007), Stabilizing greenhouse gas concentrations at low levels: an
  assessment of reduction strategies and costs, \textit{Climatic Change},
  \textit{81}, 119--159, \doi{10.1007/s10584-006-9172-9}.

\bibitem[{\textit{Velders et~al.}(2015)\textit{Velders, Fahey, Daniel,
  Andersen, and McFarland}}]{Velders2015}
Velders, G.~J., D.~W. Fahey, J.~S. Daniel, S.~O. Andersen, and M.~McFarland
  (2015), {Future atmospheric abundances and climate forcings from scenarios of
  global and regional hydrofluorocarbon (HFC) emissions}, \textit{Atmospheric
  Environment}, \textit{123}, 200--209, \doi{10.1016/j.atmosenv.2015.10.071}.

\bibitem[{\textit{Velders et~al.}(2009)\textit{Velders, Fahey, Daniel,
  McFarland, and Andersen}}]{Velders2009}
Velders, G. J.~M., D.~W. Fahey, J.~S. Daniel, M.~McFarland, and S.~O. Andersen
  (2009), {The large contribution of projected HFC emissions to future climate
  forcing}, \textit{Proceedings of the National Academy of Sciences},
  \textit{106}, 10,949--10,954, \doi{10.1073/pnas.0902817106}.

\bibitem[{\textit{Winton et~al.}(2010)\textit{Winton, Takahashi, and
  Held}}]{Winton2010}
Winton, M., K.~Takahashi, and I.~M. Held (2010), Importance of ocean heat
  uptake efficacy to transient climate change, \textit{Journal of Climate},
  \textit{23}, 2333--2344, \doi{http://dx.doi.org/10.1175/2009JCLI3139.1}.

\bibitem[{\textit{Zhang et~al.}(2011)\textit{Zhang, Wu, and Lu}}]{Zhang2011}
Zhang, H., J.~Wu, and P.~Lu (2011), A study of the radiative forcing and global
  warming potentials of hydrofluorocarbons, \textit{Journal of Quantitative
  Spectroscopy \& Radiative Transfer}, \textit{112}, 220--229,
  \doi{10.1016/j.jqsrt.2010.05.012}.

\bibitem[{\textit{Zickfeld et~al.}(2009)\textit{Zickfeld, Eby, Matthews, and
  Weaver}}]{Zickfeld2009}
Zickfeld, K., M.~Eby, H.~D. Matthews, and A.~J. Weaver (2009), Setting
  cumulative emissions targets to reduce the risk of dangerous climate change,
  \textit{Proceedings of the National Academy of Sciences of the United States
  of America}, \textit{106}, 16,129--16,134,
  \doi{http://dx.doi.org/10.1073/pnas.0805800106}.

\bibitem[{\textit{Zickfeld et~al.}(2012)\textit{Zickfeld, Arora, and
  Gillett}}]{Zickfeld2012}
Zickfeld, K., V.~K. Arora, and N.~P. Gillett (2012), Is the climate response to
  carbon emissions path dependent?, \textit{Geophysical Research Letters},
  \textit{39}, 1--6, \doi{10.1029/2011GL050205}.

\end{thebibliography}
 
\end{document}